# An Enormous Class of Double Half-Heusler Compounds with Low Thermal Conductivity

*Shashwat Anand, Max Wood, Chris Wolverton, and G. Jeffrey Snyder*[*]

Department of Materials Science and Engineering
Northwestern University
Evanston, IL 60208, US
E-mail: [*jeff.snyder@northwestern.edu](mailto:jeff.snyder@northwestern.edu).

## Abstract

Since their discovery around a century ago, the structure and chemistry of the multi-functional half-Heusler semiconductors have been studied extensively as three component systems. The elemental groups constituting these ternary compounds with the nominal formula *XYZ* are well established. From the very same set of well-known elements we explore a phase space of quaternary double ($X'X''Y_2Z_2$, $X_2Y'Y''Z_2$, and $X_2Y_2Z'Z''$), triple ($X_2'X''Y_3Z_3$) and quadruple ($X_3'X''Y_4Z_4$) half-Heusler compositions which 10 times larger in size. Using a reliable, first-principles thermodynamics methodology on a selection of 347 novel compositions, we predict 127 new stable quaternary compounds, already more than the 89 reported almost exhaustively for ternary systems. Thermoelectric performance of the state-of-the-art ternary half-Heusler compounds are limited by their intrinsically high lattice thermal conductivity ($\kappa_L$). In comparison to ternary half-Heuslers, thermal transport in double half-Heuslers is dominated by low frequency phonon modes with smaller group velocities and limited by disorder scattering. The double half-Heusler composition $Ti_2FeNiSb_2$ was synthesized and confirmed to have a significantly lower lattice thermal conductivity (factor of 3 at room temperature) than TiCoSb, thereby providing a better starting point for thermoelectric efficiency optimization. We demonstrate a dependable strategy to assist the search for low thermal conductivity half-Heuslers and point towards a huge composition space for implementing it. Our findings can be extended for systematic discovery of other large families of multi-component intermetallic semiconductors.

## Introduction

Multi-component phase spaces of intermetallic systems are mostly being explored for high-entropy metallic alloys[1–3] with impressive mechanical properties.[4,5] On the other hand, the search for complex functional semiconductors in intermetallic systems is not pursued as frequently due to (a) lack of well defined composition/ structure in a very large phase space and (b) lack of a well defined target application which could benefit from such compositional complexity. Relatively simpler intermetallic semiconductors however, such as the Half-Heusler compounds (nominal composition *XYZ*) have been known for around a century.[6] Unlike other well studied semiconductor families such as III-V, II-VI and their structural derivatives,[7] the half-Heusler compounds are constituted by a significant fraction of the elements in the periodic table (see Figure 1 a). The *X*- and *Y*-site atoms in half-Heusler structure typically belongs to the early and late transition metal groups respectively (see Figure 1 a).[6] The *Z*-site atoms belong to post transition metal p-block elements such as In, Sn, Bi etc (see Figure 1 a).[6] Thus, the chemistry of the half-Heusler structure has a lot of scope for compositional complexity.

In the last decade, the half-Heusler compounds have attracted significant research attention for their thermoelectrics properties.[8–11] Despite their chemical similarities, different compounds within this class of materials exhibit very different electrical and thermal transport properties. The best thermometric performance is observed in NbFeSb,[12] TaFeSb[13] and ZrCoBi[14] for *p*-type, and TiNiSn[10,15] for *n*-type transport due to key differences in their electronic structure. In each of these cases however, the actual high performance compositions ($Nb_{0.88}Hf_{0.12}FeSb$,[12] $Ta_{0.74}V_{0.1}Ti_{0.16}FeSb$,[13] $ZrCoBi_{0.65}Sb_{0.15}Sn_{0.20}$,[14] and $Ti_{0.5}Zr_{0.25}Hf_{0.25}NiSn_{0.998}Sn_{0.002}$[10,15] respectively) are significantly more complex than the simple *XYZ* composition primarily to reduce its large intrinsic lattice thermal conductivity (typically > 15 W/m-K at T = 300 K).[16] Such complex multi-component alloy compositions are also chosen sometimes to favorably modify electronic structure of the parent compound to enhance thermoelectric performance.[14] Hence, it is desirable to discover more complex quaternary semiconductor compounds with the same structure type and chemistry as half-Heusler compounds which form the basis of its high thermoelectric power factor,[17] but with possibly lower intrinsic lattice thermal conductivities. Further changes to such a starting composition would present a much larger phase space of quaternary compositions to aid tunability of thermoelectric properties. In addition, this class of functional semiconductors can also find use in other applications for which the ternary half-Heuslers are already being studied such as transparent conducting thin-films (e.g. TaIrGe[18]), topological semi-metals (e.g. HfIrAs[19]), and spintronics (e.g. $V_{0.8+\delta}CoSb$[20, 21]) etc.[6]

Recent discovery of defective half-Heuslers such as $Nb_{0.8}CoSb$[21–24] emerging from a Zintl chemistry[9] understanding of the structure suggests that the number of actual half-Heusler compounds available may be much larger than the ones investigated previously. Within the framework of Zintl chemistry, the half-Heusler structure can be understood as a cation $X^{n+}$ sub-lattice and an $sp^3$ covalently-bonded tetrahedral 'polyanionic' network $(YZ)^{n-}$ (see Figure 1 b). The valence on each atom (e.g. n+ for *X*-atom) of a typical semiconducting half-Heusler is assigned according to a Zintl-Klemm concept ($Ti^{+4}$, $Co^{-1}$, and $Sb^{-3}$ in TiCoSb).[9] As the atom corresponding to each of the three sites can belong to one of many elemental groups (see Figure 1 a), the corresponding valence can take multiple values.[6] The degree-of-freedom in the choice of valence allows (see Figure 1 b) for a large phase space of possible ternary compositions constrained by the valence balanced rule,[24] according to which the net valence (NV) of the three components add up to 0. Examples of valence balanced compositions are p-block based 8-

electron system LiSiAl (e.g. NV = 1 (Li$^{+1}$ s$^0$) - 4 (Si$^{-4}$ s$^2$p$^6$) + 3 (Al$^{+3}$ s$^0$p$^0$) = 0), transition metal based 18-electron system TiCoSb (e.g. NV = 4 (Ti$^{+4}$ s$^0$d$^0$) - 1 (Co$^{-1}$ d$^{10}$) - 3 (Sb$^{-3}$ s$^2$p$^6$) = 0), f-block based 27-electron DyNiSb (e.g. NV = 3 (Dy$^{+3}$ s$^0$d$^0$) + 0 (Ni$^0$ d$^{10}$) - 3 (Sb$^{-3}$ *s2p6*) = 0) and the defective Nb$_{0.8}$CoSb (NV = 4 (0.8 Nb$^{+5}$ s$^0$d$^0$) - 1 (Co$^{-1}$ d$^{10}$) - 3 (Sb$^{-3}$ s$^2$p$^6$) = 0[22, 23]). The Zintl valence balanced rule also form the basis of deterministic counting rules for ternary half-Heusler surfaces.[25] While this rule has been exploited extensively for ternary systems, it is in no way restricted to work only for three component systems.

## Results and discussion

Consider the example of the pseudoternary TiFe$_x$Co$_y$Ni$_{1-x-y}$Sb (see Figure 1 c), which allows for the aliovalent substitution of Fe, Co, and Ni atoms on the atomic *Y*−site (see Figure 1 a). The end-members, namely TiFeSb, TiCoSb, and TiPtSb all have different net valence (NV = -1, 0 and 1). In such a compositional phase space the valence balanced end-members (like TiCoSb in Figure 1 c) are candidates to form defect-free half-Heusler phases and have been studied rigorously either as experimental reports[13,19,26,27] or first-principles thermodynamic predictions[13,19,26–28]. Compositions lying close to these semiconducting end members (see cyan region in Figure 1 c) are explored routinely for studying their *p*-type[29] properties. Systems of the non-valence balanced (NV ≠ 0) end-members (TiFeSb and TiPtSb in Figure 1 c) on the other hand, were not expected to form in the Heusler phases and have attracted some attention only recently.[21–24,30] The two NV ≠ 0 systems shown (Ti-Fe-Sb and Ti-Pt-Sb) are reported to form defective Heusler phases (TiFe$_{1.5}$Sb and Ti$_{0.75}$PtSb respectively) to attain a more stable electronic configuration.[24,30] Possibly due to only one 'likely' (NV = 0) half-Heusler candidate among the three end-members of the pseudoternary, a large composition space lying in the middle have never been investigated previously (grey region in Figure 1c). As a result, the valence balanced compositions like the 'double' half-Heusler Ti$_2$FePtSb$_2$ (4 (Ti$^{+4}$ $s^0d^0$) - 1 (0.5 Fe$^{-2}$ $d^{10}$) - 0 (0.5 Pt$^0$ $d^{10}$) - 3 (Sb$^{-3}$ $s^2p^6$) = 0), and the alloy compositions connecting it to TiCoSb have largely remained ignored (see Figure 1 c).

We focus only on the quaternary half-Heusler compounds, the stability of which can be predicted from first-principles thermodynamic calculations on ordered ground state structures. Similar to Ti$_2$FePtSb$_2$, double half-Heusler candidates based on equal occupancies on the $X-$ (ScVCo$_2$Sb$_2$) and $Z-$site (Ti$_2$Ni$_2$InSb) can also be conceived. Furthermore, one can think of additional 'triple' and 'quadruple' half-Heusler compositions as well (e.g. Nb$_2$MgCo$_3$Sb$_3$ and Nb$_3$LiCo$_4$Sb$_4$) which obey the valence balanced rule (NV = 0), leading to a phase space of 7719 possible quaternary half-Heusler compositions. This number is over 10 times as large as that of the regularly studied defect-free ternary systems (715) based on the same set of elements (see Figure 1 a)! A considerable fraction of these compositions which phase separate into competing phases with alternate compositions are either (a) not contained explicitly in the inorganic crystal structure database (ICSD)[31] or (b) not examined for thermodynamic phase stability through rigorous computations[13,19,26–28]. Even excluding such cases the fraction of explored compositions for defect-free ternary systems is 69 %. In stark contrast, only 0.07 % of the possible quaternary half-Heusler compositions have been explored with 4 reports (Ti$_2$FeNiSb$_2$, ScNbNi$_2$Sn$_2$, Zr$_2$Ni$_2$InSb, and Hf$_2$Ni$_2$InSb) of the half-Heusler phase in past experiments. There are no reports – measurements or calculations – of thermal conductivity on such quaternary compositions.

We select 351 quaternary half-Heusler (315 double half-Heusler, 18 triple half-Heusler and 18 quadruple half-Heusler) compositions to perform high-throughput calculation for examining the thermodynamic stability using convex-hull analysis (see Figure 3) within the Open Quantum Materials Database (OQMD).[32,33] We restrict ourselves only to compositions bismuthide, antimonide and stannate compositions which are widely reported in the half-Heusler phase and are relatively less toxic. Furthermore, we choose compositions only based on a subset of elements (in bold font in Figure 1 a) well-known to occupy the three atomic sites of the half-Heusler structure.

The 315 double half-Heusler compositions studied here are divided into three classes (126 $X'X''Y_2Z_2$, 81 $X_2Y'Y''Z_2$ and 108 $X_2Y_2Z'Z''$ compositions) based on the site of substitution in the $XYZ$ structure (see Figure 1 a). Our calculations are in agreement with past report of thermodynamic stability of $Ti_2FeNiSb_2$, $ScNbNi_2Sn_2$, $Zr_2Ni_2InSb$, and $Hf_2Ni_2InSb$. We predict 48 $X'X''Y_2Z_2$, 36 $X_2Y'Y''Z_2$ and 28 $X_2Y_2Z'Z''$ new double half-Heusler compounds (see plus symbols in Figure 3), suggesting that substitution on either of the three atomic sites are quite favorable. Out of the 36 possible triple half-Heusler ($X_2'X''Y_3Z_3$) and quadruple half-Heusler ($X_3'X''Y_4Z_4$) compositions combined we predict 8 and 7 new compounds respectively of each sub-type (see Figure 3).

Going through the list of 127 predicted new compounds using experimental synthesis without well defined starting points can be quite daunting. To guide experimental efforts in initial choice of more likely compositions for laboratory discovery, we arrange (see color bar in Figure 3) the compounds according to their predicted stability determined by distance to the convex-Hull ($E_{CH}$). Compounds represented with darker plus symbols in Figure 3 are predicted to be more stable.

We also classify the likelihood of a predicted quaternary half-Heusler to be depending on whether there are past experimental reports of the half-Heusler phase in associated NV ≠ 0 ternary systems (for example Ti-Fe-Sb and Ti-Pt-Sb for $Ti_2FePtSb_2$) or not. In this classification (see circles in Figure 3), we assume that if a half-Heusler phase is reported for an associated NV ≠ 0 ternary (for example $Ti_{0.75+\delta}PtSb$[21,24] for $Ti_2Pt_2InSb$ or $TiFe_{1.5}Sb$[30] for $Ti_2FePdSb_2$), it is stabilized by spontaneous formation of vacancies[24] or intersitials[30] in the equiatomic *XYZ* structure. In such a case, one might expect that substitutional defects in the ternary *XYZ* structure may also be likely to stabilize the quaternary half-Heusler phase. The likelihood for the formation of the quaternary half-Heusler phase based on this rationale increases depending on whether one (see thin circles in Figure 3, for example $Ti_2FePdSb_2$) or both (see thick circles in Figure 3, for example $MgTiNi_2Sb_2$) the associated NV = 0 ternaries have a half-Heusler phase reported for them. We caution that this classification based on past experimental reports may not be complete simply because associated NV = 0 systems are not studied as rigorously (9 and 8 reported compounds for NV = -1 and NV = 1 systems respectively) as the defect-free NV = 0 systems. Furthermore, these systems also form the half-Heusler phase at defective stoichiometries which might have been missed until recently.[24] We reiterate that the classification based on past experiments is done primarily to build confidence in our predictions and to help in the choice of initial compositions for laboratory synthesis.

Despite their impressive electrical properties, the thermoelectric performance of half-Heusler materials is typically limited by an intrinsically high lattice thermal conductivity ($\kappa_L$) associated with their stiff elastic properties. Unlike the low lattice thermal conductivity of common thermoelectric materials such as $Bi_{0.4}Sb_{1.6}Te_3$ (∼ 1 W/m-K), the intrinsic lattice thermal conductivity of most well-known half-Heuslers is typically higher than > 15 W/m-K. For two materials with comparable bulk properties, such as debye temperature, average mass per unit cell, specific heat capacity and gruniesen parameter, the magnitude of lattice thermal conductivities ($\kappa_L$) depend primarily on the number of atoms in the primitive unit cell (*N*).[35] Complex materials with larger *N* have smaller $\kappa_L$ due to a relatively small fraction of high group velocity ($v_g$) acoustic modes compared to lower $v_g$ optical modes. Well-known examples of low $\kappa_L$ materials with a large $N$ are $La_2Mo_2O_9$ ($N$ = 624, $\kappa_L$ = 0.7 W/ m-K),[36] $Al_{14.7}Mn_{3.5}Si_{1.8}$ ($N$ = 138, $\kappa_L$ = 1.5 W/ m-K),[37] $Yb_{14}AlSb_{11}$ ($N$ = 104, $\kappa_L$ = 0.6 W/ m-K),[38] $LaMgAl_{11}O_{19}$ ($N$ = 64, $\kappa_L$ = 1.2 W/ m-K),[36] $LaPO_4$ ($N$ = 24, $\kappa_L$ = 2.5 W/ m-K)[36].[35] For ternary half-Heuslers however, $N$ = 3 indicating the huge potential for low thermal conductivity materials discovery if the effective $N$ can be increased systematically as in quaternary half-Heuslers.

To verify this strategy for systematic discovery of low thermal conductivity half-Heuslers, we compare measured $\kappa_L$ of double half-Heusler $Ti_2FeNiSb_2$ to that of its corresponding ternary system TiCoSb with the same average atomic mass. At room temperature, $\kappa_L$ of $Ti_2FeNiSb_2$ is smaller than TiCoSb by a factor of 3 (see Figure 4 a). To examine the origin of the smaller $\kappa_L$ in $Ti_2FeNiSb_2$ we calculate the lattice thermal conductivity of its ordered ground state from first-principles considering only three-phonon processes (see Figure 4 a). The calculated $\kappa_L$ is very similar to the measured values, especially at higher temperatures. The weaker temperature dependence in the measured $\kappa_L$ of $Ti_2FeNiSb_2$ can possibly be attributed to the presence of alloy scattering mechanism associated with disorder in the Fe/Ni sub-lattice. While no superlattice peaks associated with Fe/Ni ordering were observed in the powder XRD pattern of $Ti_2FeNiSb_2$ (see Figure 5), recent reports of short range ordering observed in electron diffraction pattern of defective ternary half-Heuslers[40] suggests that short range ordering could also exist in double half-Heusler compounds. The alloy scattering mechanism is not captured in our calculations which were performed on the ordered ground state structure of $Ti_2FeNiSb_2$.

Calculated cumulative $\kappa_L$ in TiCoSb and $Ti_2FeNiSb_2$ as a function of frequency (see Figure 4 b) suggests that over 90 % of the thermal transport in both compounds occurs in phonon modes corresponding to the acoustic modes of the ternary structure. The difference in the $\kappa_L$ between the two compounds occurs primarily in the high energy range (7-20 meV, see shaded region in Figure 4 b) corresponding to the acostic modes of TiCoSb. Both compounds largely show similar dependence of group velocity ($v_g$) as well as scattering rates on phonon frequencies (see Figure 4). However, the group velocity ($v_g$) of phonons in Ti$_2$FeNiSb$_2$ is in general much lower than the corresponding longitudinal and transverse acoustic modes of TiCoSb (see Figure 4 c). As can been understood from simple ball-spring models for cell-doubling,[35] the difference in $v_g$ becomes more prominent in the higher frequency range (10-20 meV) of the acoustic modes in TiCoSb (see Figure 4 c). The phonon scattering rates of the double half-Heusler Ti$_2$FeNiSb$_2$ structure are also slightly larger in the 7-12 meV frequency range (see Figure 4 d).

While other quaternary compounds can be investigated for thermodynamic stability, the current prediction of 127 compounds is already greater than the 89 extensively predicted /reported[13,19,26,27,31] transition metal based ternary defect-free half-Heusler phases. This result suggests that, despite a hundred years of Heusler history only a small fraction of half-Heuslers have been investigated till now. Given the intrinsically smaller thermal conductivity of quaternary half-Heusler compounds due to its complex crystal chemistry, they present a new avenue for thermoelectric materials discovery. Among the new compounds predicted here, 28 are based on relatively abundant or inexpensive elements (namely Li, Mg, Y, Ti, Zr, Hf, V, Nb, Ta, Fe, Ni, Co, Al, Ga, In, Sn, Sb and Bi) commonly used for synthesis of half-Heusler compounds, and can be immediately looked into for experimental synthesis.

The present work deals only with quaternary equivalents of half-Heusler compounds with 18-valence electrons, the concepts discussed in the present work are applicalble to other semiconducting Heusler based structures such as 8-electron Nowotnye-Juza phases (for example LiAlSi[41,42]), f -block element based half-Heuslers (for example DyNiBi[43]), 24-electron full-Heuslers (for example VFe$_2$Al[6,44,45]), and Li-based 18-electron Heuslers[46] opening up the possibility for an even larger phase space of materials with versatile properties.

In conclusion, we demonstrate that the number of known Heusler based semiconductors, despite its hundred years of history, may only be a small fraction of the actual number of stable compounds available from the same intermetallic chemistry. The previously unexplored semiconductor compositions can be conceived by choosing aliovalently substituted combinations which are Zintl valence balanced and predicted systematically using first principles thermodynamics for rapid laboratory discovery. From this strategy we discover quaternary (double, triple and quadruple) half-Heusler compounds with thermal conductivities intrinsically lower than traditional ternary half-Heuslers owing to their complex crystal chemistry. As a result, we significantly advance the decade long search for low thermal conductivity half-Heuslers without tampering with its underlying chemistry which forms the basis for their exciting properties suitable for thermoelectrics, spintronics, topological band structures and transparent conducting thin-films.

## Methods Section

***First-principles thermodynamics calculations***: The density functional theory (DFT) calculations[47] in this study were performed using Vienna ab initio simulation package (VASP).[48] We have used Perdew–Burke–Ernzerhof (PBE) formulation of the exchange–correlation energy functional derived under a gradient-generalized approximation (GGA).[49] Plane-wave basis sets are truncated at energy cutoff of 315 eV, and a Γ-centered k-point meshes with a density of 8000 k-points per reciprocal atom (KPPRA) was used. All structures were relaxed with respect to cell vectors and their internal degrees of freedom until forces on all atoms were less than 0.1 eV nm$^{-1}$ before calculating their formation enthalpies ($E_{form}$). The Open Quantum Materials Database (OQMD)[32,33] was used for the convex hull construction to check for thermodynamic stability against all other known phases in the respective composition space. Lattice thermal conductivity calculations were performed assuming only the presence of three-phonon processes within an Umpklapp scattering mechanism.[50, 51]

***Synthesis and thermal conductivity measurements of double half-Heusler***: The sample of Ti$_2$FeNiSb$_2$ was prepared using an arc melter using stoichiometric ratios of bulk Ti slugs (99.99% Sigma-Aldrich), Fe lumps (99.99% Alpha Aesar), Ni slugs (99.99% Alpha Aesar), and Sb shot (99.999% Alpha Aesar). Starting elements were cut into small pieces and loaded into an edmund-buehler MAM-1 arc melter where they were melted together 5 times flipping in between each melt. The arc melted button was then pulverized using a in a stainless steel vial for one hour using a SPEX Sample Prep 800 Series Mixer/Mill. This powder was consolidated using an induction heated rapid hot-press under a flowing argon atmosphere within a 1/2 in diameter high-density graphite die (POCO). This sample was then pressed at 1100 /textdegree C for one hour at a pressure of 45 MPa. Once pressed this sample was polished to remove excess graphite from its surface, sealed in an evacuated fused silca ampule, and annealed at 900 /textdegree C for one week.

Transmission XRD was performed on powder pulverized from the pellet using a Stoe STADI-MP x-ray diffractometer using CuKα radiation. Electronic transport measurements were measured under dynamic high vacuum up to 875 K with a ramp speed of 75 K/h. Resistivity and hall coefficient were measured concurrently using the Van Der Pauw technique with pressure-assisted molybdenum contacts equipped a 2 T magnet41. Thermal diffusivity measurements were taken with a Netzch LFA 457 under purged flowing argon up to 875K with a ramp speed of 75 K/h. Thermal conductivity was calculated estimating heat capacity with the Dulong-Petit law. Seebeck coefficient was measured under dynamic high vacuum up to 875 K with a homebuilt system using Chromel-Nb thermocouples 42. Lattice Thermal Conductivity was calculated by subtracting the electronic contribution from the calculated total thermal conductivity. The electronic portion of thermal conductivity was found using the Weidemann-Franz law, in which a variable Lorenz number was used based on the sample's Seebeck coefficient.

**Acknowledgements**
SA would like to thank Vinay Ishwar Hegde for fruitful discussion on the subject of thermodynamic phase stability and high throughput calculations. G. J. S and S. A acknowledge support by the "Designing Materials to Revolutionize and Engineer our Future" program of the National Science Foundation, under Award No. 1729487 and the U.S. Department of Energy, Office of Energy Efficiency and Renewable Energy (EERE). C.W. (DFT calculations) acknowledges support by the U.S. Department of Energy, Office of Science and Office of Basic Energy Sciences, under Award No.DE-SC0014520.

# Figures

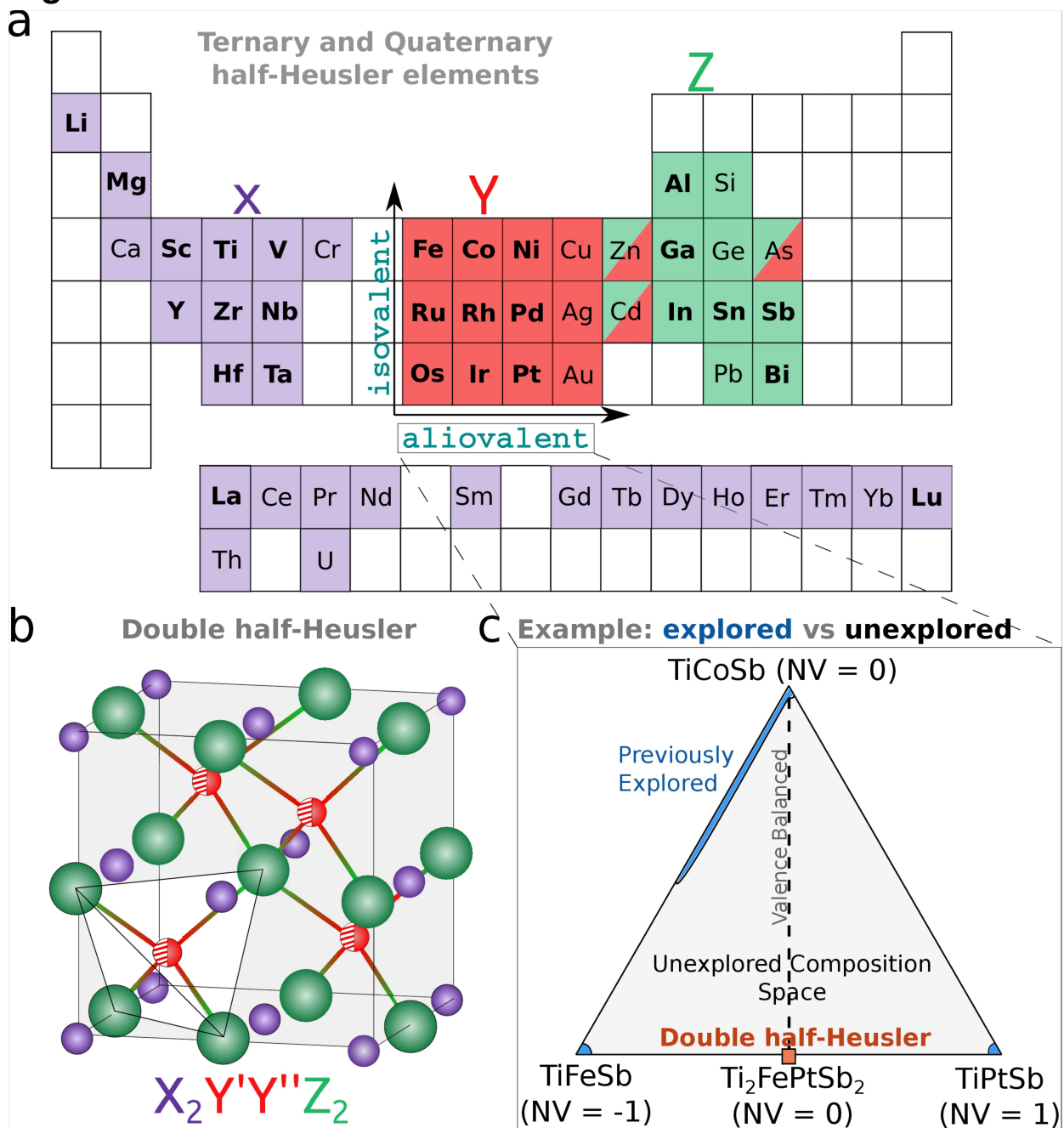

**Figure 1.** (a) Color scheme in the periodic table representing elements occupying various sites (*X* (violet), *Y* (red) and *Z* (green)) of the cubic half-Heusler structure (in panel b). Quaternary half-Heusler compositions selected in the present work are based on elements given in bold font. (b) The double half-Heusler structure with the general formula $X_2Y'Y''Z_2$ has equal occupancy on the *Y*-site (in half red/ half red stripes) such that the overall composition is valence balanced (NV = 0). The covalently bonded 'polyanionic' sublattice $(YZ)^{n-}$ is represented by the unshaded tetrahedron. (c) Current status of explored compositions (blue) in the TiFe$_x$Co$_y$Pt$_{1-y}$Sb example pseudoternary based on aliovalent substitution on the *Y*-site. Double half-Heusler Ti$_2$FePtSb$_2$ (orange square) and the compositions joining it to TiCoSb are valence balanced (NV = 0) and have not been investigated previously (grey).

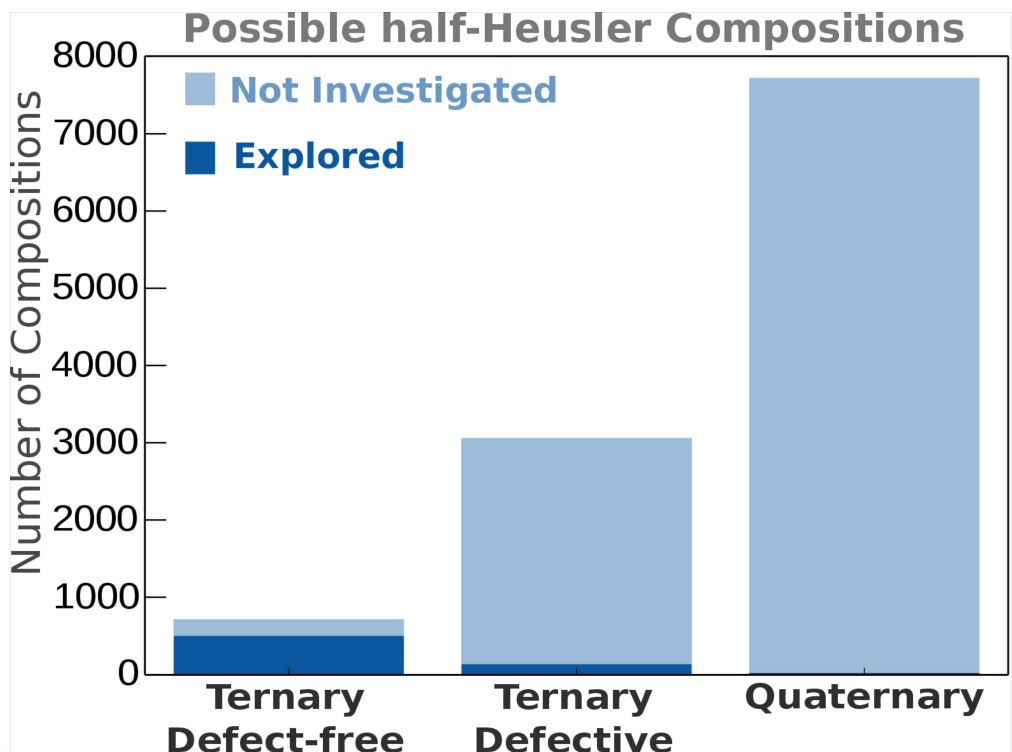

**Figure 2.**
Bar chart depicting the current status of exploration in possible ternary half-Heusler systems as opposed to quaternary systems. Compositions were obtained by imposing the valence balanced rule on the elemental combinations provided in figure 1 a. The dark blue color gives a conservative estimate of the fraction of explored compositions. The quaternary phase space is almost completely unexplored.

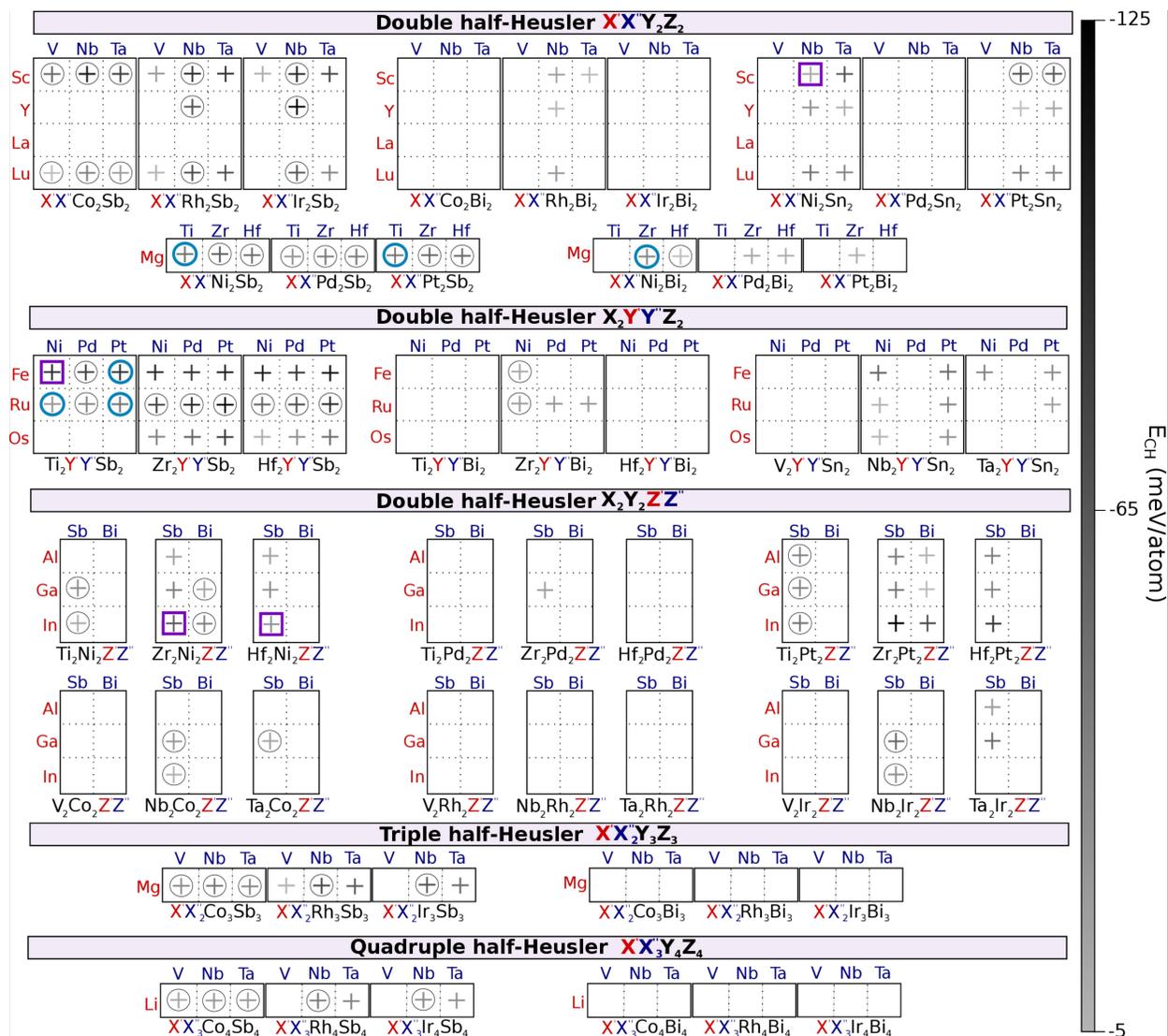

**Figure 3.** 351 quaternary half-Heusler compositions (divided between 5 sub-types) investigated in the present work for stability in the half-Heusler structure. Half-Heuslers predicted stable are represented by a plus (+) symbol. For compounds where the half-Heusler phase is predicted unstable, the space is left blank. The color bar on the side gives the magnitude of predicted stability ($E_{CH}$, convex-hull distance) of the quaternary half-Heusler. Predicted compounds for which one of the associated NV = 0 ternary systems have a half-Heusler phase reported experimentally are denoted by thin circles. Predicted compounds for which one of the associated NV = 0 ternary systems have a half-Heusler phase reported experimentally are denoted by thick circles. Previously reported quaternary half-Heuslers are denoted by boxes.

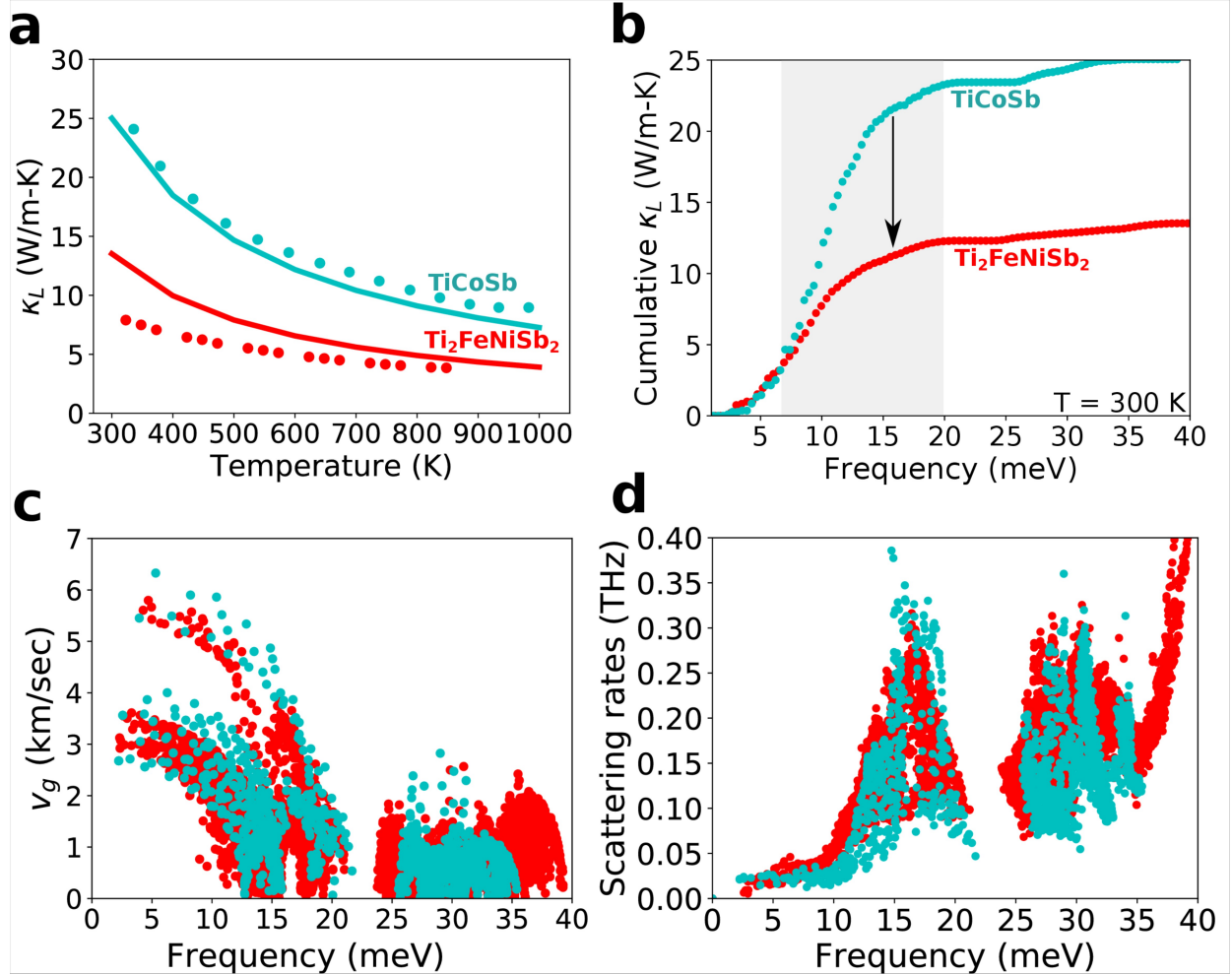

**Figure 4.** (a) Calculated (line) and measured (scatter points) lattice thermal conductivity ($\kappa_L$) of TiCoSb (cyan) and Ti$_2$FeNiSb$_2$ (red) as a function of temperature. Experimental values of TiCoSb are taken from Sekimoto et al.[39]. Calculated phonon frequency dependence of (b) cumulative $\kappa_L$, (c) group velocities ($v_g$), and (d) phonon-phonon scattering rates in the two compounds at T = 300 K.

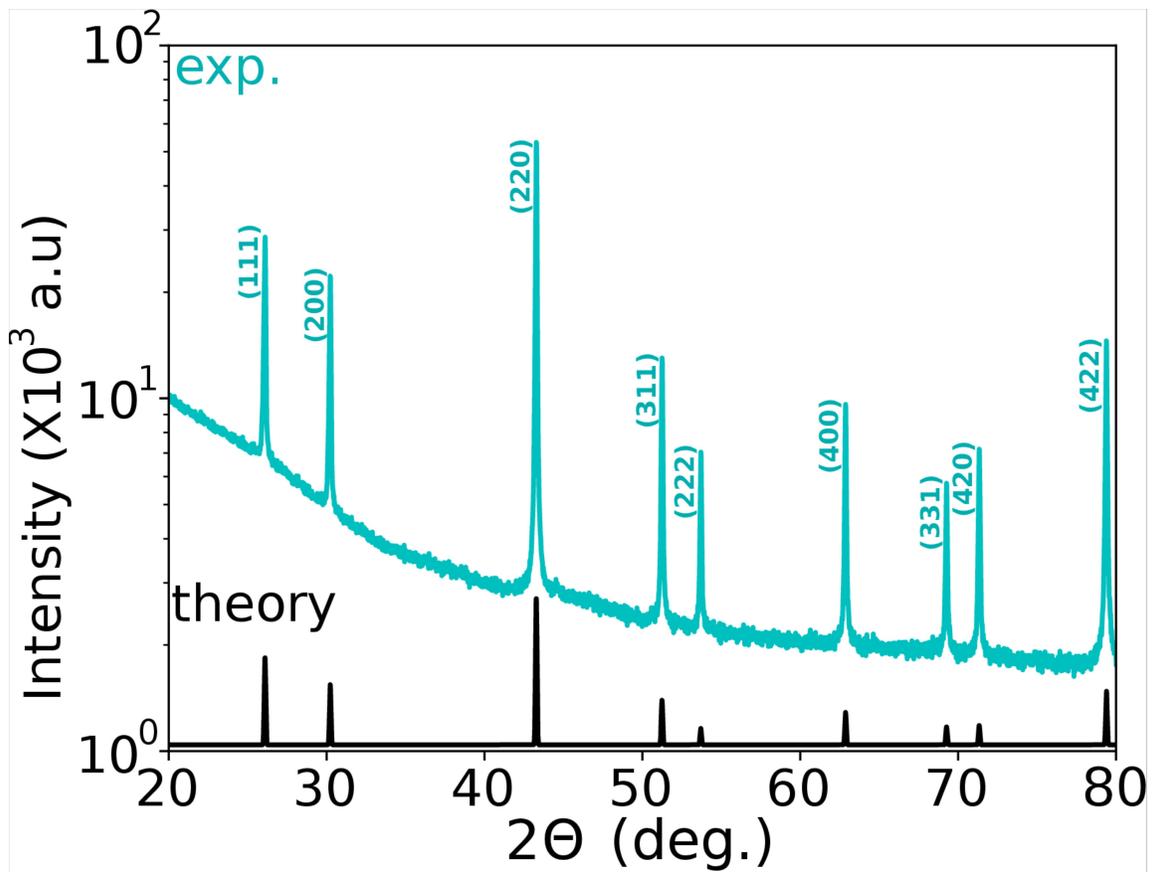

**Figure 5.** Room temperature powder XRD pattern (cyan) of $Ti_2FeNiSb_2$ annealed at 1073 K. Theoretical XRD pattern of special quasi-random structure (SQS) of $Ti_2FeNiSb_2$ is given in black.